# Tuning magnetic and optical properties through strain in epitaxial LaCrO$_3$ thin films

Yogesh Sharma[1,*], Binod Paudel[1,2], Jegon Lee[3], Woo Seok Choi[3], Zhenzhong Yang[4], Han Wang[4], Yingge Du[4], Kyeong Tae Kang[1], Ghanshyam Pilania[5], and Aiping Chen[1]

[1]Center for Integrated Nanotechnologies (CINT), Los Alamos National Laboratory, Los Alamos, NM 87545, USA
[2]Department of Physics, New Mexico State University, Las Cruces, New Mexico 88003, US
[3]Department of Physics, Sungkyunkwan University, Suwon 16419, South Korea.
[4]Physical Sciences Division, Physical and Computational Sciences Directorate, Pacific Northwest National Laboratory, Richland, Washington 99352, USA
[5]Materials Science and Technology Division, Los Alamos National Laboratory, Los Alamos, NM 87545, USA

*ysharma@lanl.gov

**ABSTRACT**

We report on the effect of epitaxial strain on magnetic and optical properties of perovskite LaCrO$_3$ (LCO) single crystal thin films. Epitaxial LCO thin films are grown by pulsed laser deposition on proper choice of substrates to impose different strain states. A combined experimental and theoretical approach is used to demonstrate the direct correlation between lattice-strain and functional properties. The magnetization results show that the lattice anisotropy plays a critical role in controlling the magnetic behavior of LCO films. The strain induced tetragonality in the film lattice strongly affects the optical transitions and charge transfer gap in LCO. This study opens new possibilities to tailoring the functional properties of LCO and related materials by strain engineering in epitaxial growth.

## I. INTRODUCTION

The perovskite rare-earth orthochromites ($RCrO_3$, where R = rare earth elements) show a broad range of functional properties such as, magnetic exchange bias [1–3], anisotropic magnetostriction [4], magnetoelectric-multiferroicity [3,5–9], magnetocaloric effect [10–12], photocatalysis [13,14], and relaxor dielectric behavior [15–17]. Such a wide range of functionalities make them interesting materials for both the fundamental physical studies and applied research. The $RCrO_3$ perovskites crystallize in the centrosymmetric orthorhombic (*Pbnm*, $GdFeO_3$-type) structure with four formula units per unit cell and the corner sharing of $CrO_6$ octahedra through oxygen [18,19]. The hybridization between O $2p$ and chromium $3d$ ($Cr^{3+}$ $3d^3$ configuration) orbitals results in the multiple optical absorptions within a wide spectral range from ultraviolet to visible light [20,21]. These isostructural $RCrO_3$ perovskites show a G-type antiferromagnetic (G-AFM) spin order, where the dominant magnetic contribution stems from the antisymmetric $Cr^{3+}$–O–$Cr^{3+}$ AFM superexchange interaction via the hybridization [18,22–25]. Due to the antisymmetric superexchange between $Cr^{+3}$ spins, a weak ferromagnetism (WFM) arises below the Néel temperature ($T_N$) from a slight canting of the AFM spins that lie either along the *a*- or *c*-axis of the unit cell [1,22]. The complex exchange interaction leads a strong coupling of the spin and lattice degrees of freedom, manifesting the magnetostructural coupling underlying the observed polar order in some $RCrO_3$ perovskites [7,9].

Recent observations on tuning the structure, magnetic and optical properties of bulk $RCrO_3$ by chemical-strain, misfit-strain and hydrostatic pressure have attracted significant attention towards the functional applications of this class of materials [26–31]. The lattice structure, net magnetic moment, magnetic ordering temperature, and optical band-gap can be tuned through the change of internal strain in $RCrO_3$ lattice, either by external applied pressure or chemical strain [21,26,27,31]. Enhanced magnetization and reduced optical band gap are observed due to the application of hydrostatic pressure and chemical strain, respectively [27,28,30]. Theoretical studies reports that the epitaxial-strain modifies antipolar displacements, Cr–O–Cr bond angles, Cr–O bond lengths and oxygen octahedral tilt within the lattice symmetry (*Pbnm*), which significantly affect the magnetic behavior of $RCrO_3$ [26]. The theory results also suggest that the optical band gap of $RCrO_3$ can be decreased by as large as 0.2 eV by judiciously choosing the strain conditions [21]. However, despite the theoretical predictions, available experimental evidences pertaining to the effect of epitaxial strain on the magnetic and optical properties of $RCrO_3$ are highly limited [17,28,32]. Direct experimental observation of strain-tunability of the $RCrO_3$ perovskites can offer new opportunities to tailor their physical properties and expand practicality for various functional applications.

Here, we perform experimental and theoretical studies to reveal the effect of strain on the magnetic and optical properties of single crystal LCO epitaxial films. We investigate the magnetic behavior and optical transitions of differently strained LCO thin films grown by pulsed laser deposition. The magnetization results show that the lattice-strain can be used to modify the magnetic anisotropy and the net magnetic moment in LCO films. The magnetic easy axis is along the in-plane direction for the tensile-strained films whereas, compressive strain results in a magnetic easy axis along the out-of-plane direction. The optical results evident that the direction of lattice tetragonality has a strong influence on the electronic structure, and hence, the optical band gap of LCO films.

**II. EXPERIMENTAL AND THEORETICAL DETAILS**

A polycrystalline ceramic target of stoichiometric $LaCrO_3$ (LCO) was synthesized using the conventional solid-state reaction method. LCO films were deposited on (001) $SrTiO_3$ (STO) and

(001) [(LaAlO$_3$)$_{0.3}$(SrAl$_{0.5}$Ta$_{0.5}$O$_3$)$_{0.7}$] (LSAT) substrates using pulsed laser deposition (PLD). A KrF excimer laser (λ = 248 nm) operating at a repetition rate of 5 Hz with a laser fluence of 1.25 J/cm$^2$ was used for target ablation. The detailed growth conditions were described elsewhere [17]. The crystal structure and growth orientation of the films were characterized by X-ray diffraction (XRD) using a Panalytical X'Pert Pro four-circle high resolution X-ray diffractometer with Cu Kα1 radiation. The microstructure of the films and interface quality between film and substrate were studied by an aberration-corrected Titan 80-300™ scanning transmission electron microscope (STEM). Cross-sectional specimens oriented along the [100] STO/LSAT direction for STEM analysis were prepared using ion milling after mechanical thinning and precision polishing. Magnetization as a function of temperature and applied magnetic field, M(T) and M(H), respectively, were measured in a physical property measurement system (PPMS, Quantum Design) with a vibrating sample magnetometer (VSM). The M(H) curves were collected in a zero-field-cooled (ZFC) setting with magnetic field applied in-plane and out-of-plane of the samples at 10K. The M(T) curves were measured in a field-cooled (FC) setting while warming the samples from 10 K to 350 K with a magnetic field of 100 Oe applied in-plane of the samples. The spectroscopic ellipsometry (M-2000, J. A. Woollam Co.) was used to obtain optical conductivity as a function of photon energy at room temperature.

Theoretical calculations were performed within the density functional theory (DFT) approach using the Vienna Ab initio Simulation Package (VASP) [33,34]. A plane-wave basis set and the generalized gradient approximation (GGA) [35] of Perdew–Burke–Ernzerhof (PBE) [36] with projector augmented wave (PAW) method [37] was used to obtain the structural, energetics and electronic properties of LCO. A plane wave cut-off energy of 520 eV was used for the plane-wave basis set and a Γ-centered Monkhorst–Pack $k$-point mesh [38] of 6 × 6 × 5 was used for the Brillouin zone sampling of the 20-atom orthorhombic cell (*Pbnm* space group) which can be viewed as a √2 × √2 × 2 multiple of the prototypical five-atom pseudo-cubic perovskite unit cell. The O 2*s* and 2*p*, the Cr 3*p*, 4*s*, 3*d*, and La 5*s*, 5*p*, 5*d*, and 6*s* electrons were included in the valence states. A 0.05 eV Gaussian smearing was applied to band occupancies near the Fermi level, and total energies were extrapolated back to 0 K. Non-collinear magnetic states were explicitly considered in the calculations and a 'Hubbard U' correction (with Dudarev's formulation; J = 0 eV and U = 3.7) [39] was employed to correctly represent the localized electronic states of Cr's 3*d* electrons [40]. To obtain a geometry optimized equilibrium structure, starting with a G-type antiferromagnetic (AFM) configuration all atomic positions, as well as the three lattice parameters were fully relaxed using the conjugate gradient method until all the Hellmann-Feynman forces and the stress component were less than 0.005 eV/Å and $1.0\times10^{-3}$ GPa, respectively. The magnetic anisotropy energy [41] for several local Cr spin alignments ranging between the IP [100] and the OOP [001] directions was considered for the relaxed and 1% IP compressive and tensile strains. For the latter two cases, the *a* and *b* lattice parameters were constrained to the respective strained values relative to the relaxed case, while the *c* lattice parameters and all the internal coordinates were allowed to relax. Spin-orbit coupling was considered in the magnetic anisotropy energy computations.

**III. RESULTS AND DISCUSSION**

Bulk orthorhombic perovskite LCO have the lattice parameters of *a* = 5.513 Å, *b* = 5.476 Å, and *c* = 7.759 Å, where the pseudocubic unit-cell within the orthorhombic structure yields $a_{pc}$ = 3.885 Å [42]. The misfit lattice-strain between LCO and STO and LCO and LSAT is 0.51% and −0.44%, respectively (Fig. 1(a)). Figures 1(a) and (b) show XRD symmetric *θ*-2*θ* scans for LCO film deposited on STO and LSAT substrates. Only 00*l* reflections of the LCO films are observed in the

θ-2θ scan, indicating that the film is epitaxial and phase pure. The LCO films of 39 nm (on STO) and 36 nm (on LSAT) exhibit well-defined thickness fringes as shown in the fine θ-2θ scans around the LCO 002 Bragg peaks in Fig. 1(b). Rocking curve measurements around the 002 reflection of the LCO films and the 002 reflection of the STO and LSAT substrates are shown superimposed in Fig. 1(c). The full width at half maximum (FWHM) of LCO films' rocking curves were measured to be ~0.04° (on STO) and ~0.06° (on LSAT). These values are close to the FWHM of their respective substrates (~0.02° for STO and LSAT), indicating the high structural quality of the films. The X-ray φ-scans were measured to further characterize the structural perfection of these films, where the in-plane epitaxial orientation relationship is determined to be (001) LCO || (001) STO/LSAT and [100] LCO || [100] STO/LSAT [43].

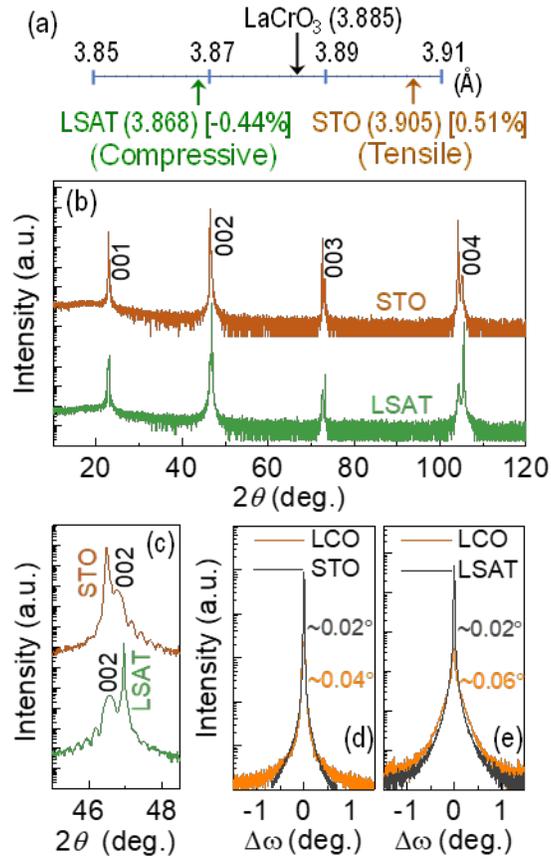

FIG. 1. (a) Lattice mismatch between pseudocubic LCO and STO and LSAT substrates. (b) XRD θ-2θ scans of the LCO films grown on STO and LSAT substrates. (c) An enlarged view of the θ-

$2\theta$ scans around the LCO 002 peak with clear thickness fringes. Rocking curves of the substrates and LCO 002 peaks are plotted together for the LCO films on (d) STO and (e) LSAT substrates.

To inspect the strain state of the LCO films, X-ray reciprocal space mapping (RSM) were performed around the asymmetric 103 Bragg's peaks of LCO film and STO and LSAT substrates, as shown in Figs. 2(a) and 2(b). The perfect vertical alignment of 103 films' peak position with respect to the substrates' peak for both samples, indicates that the films are coherently strained to the underlying STO and LSAT substrates. From RSMs, the in-plane (IP) and out-of-plane (OOP) lattice parameters are determined as $a_\parallel$ = 3.905 Å and $a_\perp$ = 3.872 Å for the LCO film on STO and $a_\parallel$ = 3.868 Å and $a_\perp$ = 3.902 Å for the LCO film on LSAT, respectively. Films' IP strain $\varepsilon_{xx}$ = 0.51% (on STO) and −0.44% (on LSAT) and OOP strain $\varepsilon_{zz}$ = −0.33% (on STO) and 0.44% (on LSAT) are calculated using $\varepsilon_{xx} = (a_\parallel - a_{pc})/a_{pc}$ and $\varepsilon_{zz} = (a_\perp - a_{pc})/a_{pc}$. The LCO film on STO is under IP tensile-strain whereas the LCO film on LSAT is under compressive-strain along IP direction. The values of Poisson ratio [$v = \varepsilon_{zz}/(\varepsilon_{zz} - 2\varepsilon_{xx})$] of the LCO films are calculated to be 0.24 (on STO) and 0.33 (on LSAT); the former is close to the values of 0.23 and 0.21 reported for epitaxial LCO and YCrO$_3$ films on STO, respectively [17,42]. The microstructure of LCO/STO (001) and LCO/LSAT (001) films are investigated using scanning transmission electron microscopy (STEM). The high-angle annular dark field (HAADF) STEM micrographs acquired along the STO/LSAT [001] zone axis, near the film/substrate interfaces are presented in Figs. 2(c-f). The microstructures of the films and the film-substrates interfaces show a single crystalline lattice with high epitaxial quality and the abrupt and fully coherent interfaces with the substrate. The STEM observations are consistent with the XRD findings that the films are structurally uniform and epitaxial.

To understand how the lattice-strain impacts the magnetic behavior, the magnetic properties of the compressive and tensile strained LCO films are examined. The M(T) curves measured in the FC and ZFC settings show the magnetic ordering onset with a $T_N$ of ~286 K for both of the films [43], which is close to the $T_N$ of ~290 K for the bulk powders and polycrystalline LCO films [8,44]. Figure 3 shows the M(H) loops of both the LCO films on STO and LSAT measured at 10 K by applying the field in OOP and IP directions of the samples. Figure 3(a) shows the IP and OOP M(H) loops of the tensile-strained LCO film on STO, where the diversity in magnetization along the two different directions and almost equal value of saturation magnetization can be observed. It shows that the easy axis is orientated along the IP direction in tensile-strained LCO film on STO (Fig. 3(a)). A different magnetic behavior was observed in the case of compressive-strained LCO film on LSAT, where the field direction dependence of magnetization is reversed and the easy axis of magnetization is along the OOP direction as shown in Fig. 3(b). A relatively stronger hysteresis (right inset of Fig. 3(b)) and a higher saturation magnetic moment are observed in the IP and OOP M(H) loops of the LCO film on LSAT (Fig. 3(b)). The magnetization results show that the epitaxial strain dependence of the lattice anisotropy can be used to modify magnetic behavior in LCO thin films.

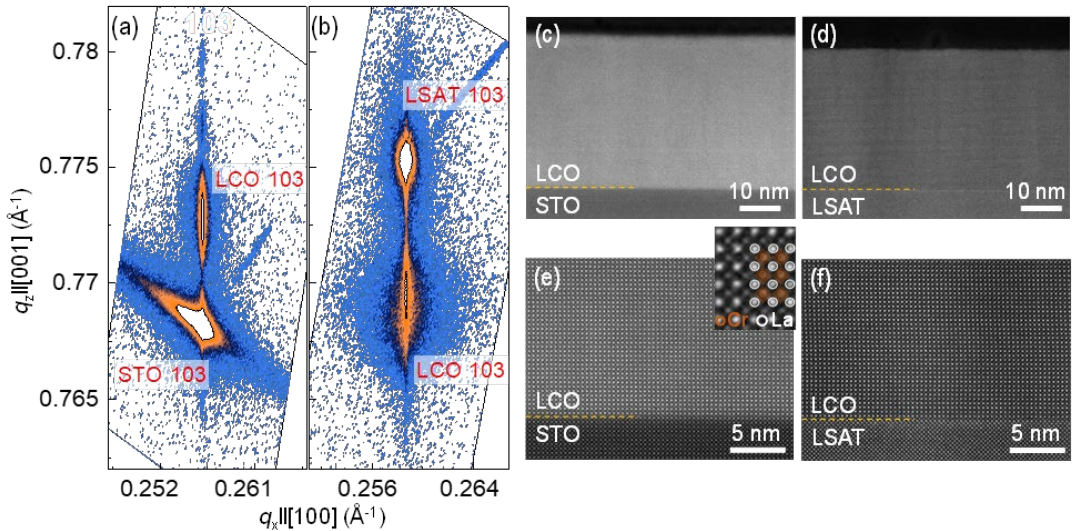

FIG. 2. RSMs around the 103 reflections of the LCO and substrates for the films on (a) STO and (b) LSAT. RSMs were recorded on the same samples characterized in Fig. 1. HAADF-STEM images with a low magnification showing the microstructure of the same samples, the LCO films

on (c) STO and (d) LSAT substrates, characterized by RSM. The magnified cross-sectional HAADF-STEM images for the LCO films on (e) STO and (f) LSAT substrates. The STEM observations are consistent with the X-ray findings that both the films show high epitaxial quality and fully coherent interfaces with the underlying substrates.

To understand the role of lattice-strain on modulating magnetic anisotropy in LCO films, we resort to DFT computations. The magnetic anisotropy energy (MAE) is calculated for different spin-orientation angles ($\theta$), where local spins on the Cr sublattice are gradually varied between IP [100] and OOP [001] directions for LCO in the relaxed as well as with 1% IP compressive and tensile strains, as shown in Fig. 4. For the relaxed case, the most stable magnetic configuration is the one where the local spins on the Cr ions are aligned very close to the IP [100] direction (although not exactly due to the slightly canted G-AFM ground state of LCO). As the local spins on the Cr atoms are moved away from the easy magnetization axis [100] along the out of plane axis [001] by gradually increasing the angle $\theta$, the MAE increases indicating that the easy axis is oriented along the IP direction. For the 1% IP tensile strain, a qualitatively similar trend is predicted, although the anisotropy is significantly enhanced as magnetization along [001] becomes more energetically unfavored with respect to the relaxed case. On the other hand, a compressive strain of 1% leads to a drastic lowering in the MAE along the OOP [001] axis (~40 μeV/f.u. to ~7 μeV/f.u.), making this direction comparable to that of the easy axis of polarization. We note here that the energy differences of a few μeV are within the error bars of the DFT computations.

It is pertinent to note that for the bulk canted G-AFM structure the easy magnetization axis (or the direction of net magnetization) is nearly perpendicular to the microscopic local magnetic moments on the individual Cr atoms. A detailed analysis of the ground state magnetic configuration showing this, for a zero external field and zero epitaxial strain, is provided in the supporting information [43]. In presence of an external magnetic field the magnitude and direction of the net magnetization is determined by both the induced magnetization and the inherent magnetization due to the canting (if slight canting still persists).

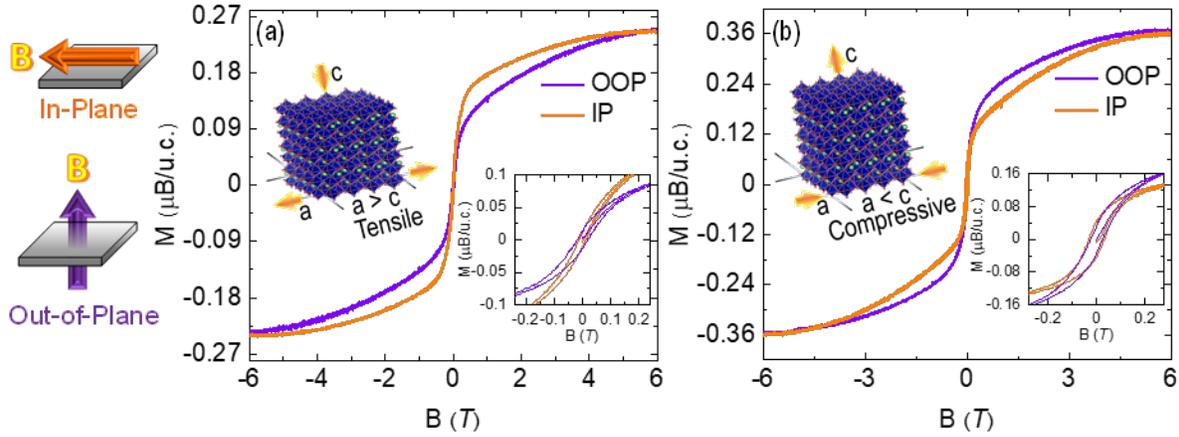

FIG. 3. M(H) loops at 10 K for the (a) LCO film on STO and (b) LCO film on LSAT. The illustrations on the left show M(H) loops were carried out by applying the magnetic field IP and OOP directions to the samples. The right insets in both (a) and (b) show enlarged views of M(H) loops. The left insets in (a) and (b) are the schematic representation of lattice mismatch-induced deformation directions on the films with the underlying substrates. The M(H) loops on the differently strained LCO films show that the lattice strain has a strong influence on the magnetic easy axis direction and overall magnetization.

We find that since the magnitude of the magnetization due to canting is about 10 to 20 times smaller than the induced saturation magnetization, for both the films with tensile and compressive epitaxial strains, and therefore the direction of the net magnetization does not remain perpendicular to the microscopic magnetization direction. Based on these observations, we assume that in the case of IP magnetic field all the local moments leading to saturation magnetization will be aligned along the [100] or near [100] direction, while for an OOP magnetic field they are aligned along the [001] or near [100] direction. Given this scenario, based on the MAE calculations we further conclude that for epitaxial tensile strain the extent of magnetization for an IP magnetic field should be relatively larger, since at $\theta = 0°$ MAE is lower than that for $\theta = 90°$. Going to an epitaxial compressive strain, however, this tendency is largely altered and MAE at $\theta = 90°$ is greatly reduced to almost 10% of its value for tensile strain. It depicts that the compressive strain largely annihilates strong tendency for magnetization along the IP direction exhibited for the tensile and relaxed cases, and might even favor a higher value of magnetization for OOP applied magnetic field, which is in agreement with our experimental measurements presented in Figure 3.

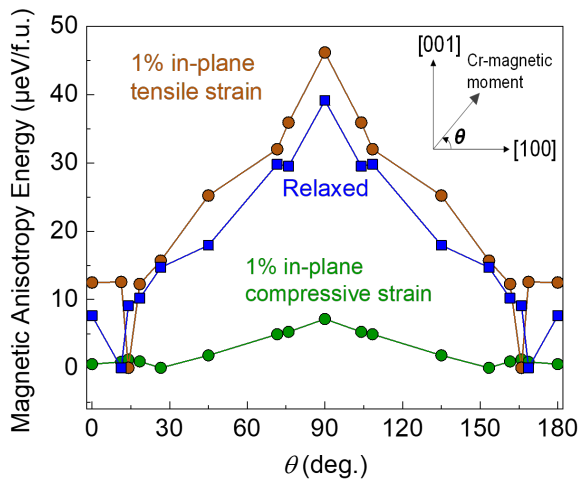

FIG. 4. The calculation of magnetic anisotropy energy for different spin-orientation angles ($\theta$), where local spins on the Cr sublattice are gradually varied between IP [100] and OOP [001] directions for LCO in the relaxed, tensile and compressive strain states.

To investigate how the lattice strain impacts the electronic structure of epitaxial LCO films, the optical properties of LCO films were studied using spectroscopic ellipsometry. Figures 5 (a) and 5(b) show the real part of the optical conductivity spectra as a function of photon energy ($\sigma_1(\omega)$) for the LCO films on STO and LSAT, respectively. The spectra in Figs. 5(a) and 5(b) are well-fitted using Lorentzian functions of $E_1$, $E_2$, and $E_3$. The characteristic energy scale of each oscillator correspond to the optical transitions between the orbital states of LCO, as shown in Fig. 5(c). In particular, $E_1$ corresponds to the $d$-$d$ transition between the occupied $t_{2g}$ to the unoccupied $e_g$ states of Cr $3d$ orbitals. Note that the on-site $t_{2g}$ to $t_{2g}$ transition is not plausible, because it involves the spin-flip transition. $E_2$ and $E_3$ correspond to charge transfer transitions from O $2p$ state to Cr $3d$ $t_{2g}$ and $e_g$ states, respectively. Our density of states calculation supports the assignment of the optical transition to the band structure as shown in Figs. S3-S5 [43]. Among the three characteristic transitions, the lowest-lying $E_1$ determines the optical band gap of the system. The directions of lattice tetragonality strongly influences the band structure, and hence, the band gap of LCO [21]. Figure 5(d) shows the atomic picture of the occupied Cr 3d $t_{2g}$ state for the simple understanding of the band gap tuning, in which the crystal field splitting of the $t_{2g}$ state owing to the strain is schematically presented. Because of the direction of the tetragonality, the energy level of the highest occupied $t_{2g}$ state under compressive strain (xy orbital) moves further up than those under tensile strain (yz/zx orbitals), effectively reducing the band gap. As marked in Figs. 5(a) and 5(b), the band gaps of the LCO film on STO and LSAT are 3.65 and 3.39 eV, respectively, consistently following the trend. The strain-induced modifications in the optical transitions and the observation of higher optical bang gap in the compressively strained LCO film on LSAT suggest the optical band gap sensitivity to the lattice strain in isostructural $RCrO_3$ perovskites.

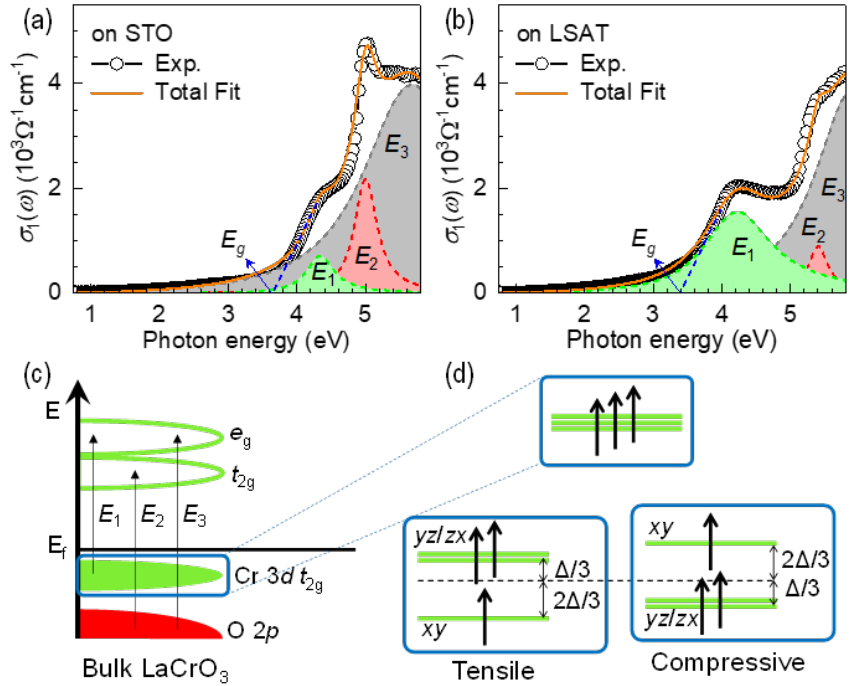

FIG. 5. Electronic structure and optical band gap of LCO thin films. $\sigma_1(\omega)$ of LCO films on (a) STO and (b) LSAT substrates, where the spectra are deconvoluted using the Lorentzian transitions indexed as $E_1$ (green), $E_2$ (red), and $E_3$ (dark-gray). The band gap ($E_g$) determined by $E_1$ is also marked with arrows. (c) The schematic band structure with the plausible optical transitions observed from the experiment. (d) Atomic energy scale of the occupied Cr 3d $t_{2g}$ level under tensile and compressive strain.

## IV. CONCLUSION

In summary, we have shown that the magnetic and optical properties of perovskite LCO thin films can be tuned by lattice strain. The highly crystalline, phase-pure epitaxial LCO thin films were grown on selected substrates to achieve different strain states. X-ray diffraction and scanning transmission electron microscopy results demonstrate excellent crystalline quality and smooth interfaces of the fully strained LCO films. Field-dependent magnetometry shows that the lattice-strain strongly affects the magnetic anisotropy and the net magnetic moment. The optical results suggest that the direction of lattice tetragonality has a strong influence on the band gap of LCO films. The strain-tunability of the magnetization and optical bang gap make LCO a promising material for multifunctional applications. Furthermore, understanding the effects of strain on functional properties of perovskite $RCrO_3$ will accelerate both fundamental studies of physical phenomena as well as the utilization of these materials in functional applications.

## ACKNOWLEDGEMENT

The work at Los Alamos National Laboratory was supported by the NNSA's Laboratory Directed Research and Development (LDRD) Program and was performed, in part, at the Center for Integrated Nanotechnologies, an Office of Science User Facility operated for the U.S. Department of Energy Office of Science. Los Alamos National Laboratory, an affirmative action equal opportunity employer, is managed by Triad National Security, LLC for the U.S. Department of Energy's NNSA, under contract 89233218CNA000001. Y.S. acknowledges the support from the LDRD program of Los Alamos National Laboratory and the G. T. Seaborg Institute under project number 20210527CR. G.P. was supported by the Los Alamos National Laboratory's LDRD program's Directed Research (DR) project number 20200104DR. J.L. and W.S.C. acknowledge support from the Basic Science Research Program through the National Research Foundation of Korea (Grant No. NRF-2019R1A2B5B02004546). Computational support for this work was provided by Los Alamos National Laboratory's high performance computing clusters. TEM and data analysis performed by Z.Y., H.W and Y.D. were supported by the U.S. Department of Energy, Office of Science, Office of Basic Energy Sciences, Early Career Research Program under award #68278 using EMSL, a DOE User Facility sponsored by the Office of Biological and Environmental Research and located at the Pacific Northwest National Laboratory.